\documentclass[aps,prl,reprint,superscriptaddress]{revtex4-1}

\usepackage{mathptmx,graphicx,amsmath,pifont,epstopdf,array,makecell,longtable}
\usepackage[T1]{fontenc}
\usepackage[colorlinks,linkcolor=blue, urlcolor=blue, anchorcolor=blue, citecolor=blue]{hyperref}

\makeatletter

\newcommand{\Rmnum}[1]{\expandafter\@slowromancap\romannumeral #1@}
\makeatother

\begin{document}	
	\title{Phase Diagram of growth modes in Graphene Growth on Cooper by Vapor Deposition}
	\author{Tongtong Wang}
	\affiliation{National Laboratory of Solid State Microstructures and Department of Physics, Nanjing University, Nanjing 210093, China}
	\affiliation{Collaborative Innovation Center of Advanced Microstructures, Nanjing University, Nanjing 210093, China}
	\author{Jian Zheng}
	\affiliation{National Laboratory of Solid State Microstructures and Department of Physics, Nanjing University, Nanjing 210093, China}
	\affiliation{Collaborative Innovation Center of Advanced Microstructures, Nanjing University, Nanjing 210093, China}
	\author{Xin Wei}
	\affiliation{National Laboratory of Solid State Microstructures and Department of Physics, Nanjing University, Nanjing 210093, China}
	\affiliation{Collaborative Innovation Center of Advanced Microstructures, Nanjing University, Nanjing 210093, China}
	\author{Dajun Shu}\email{djshu@nju.edu.cn}
	\affiliation{National Laboratory of Solid State Microstructures and Department of Physics, Nanjing University, Nanjing 210093, China}
	\affiliation{Collaborative Innovation Center of Advanced Microstructures, Nanjing University, Nanjing 210093, China}	
	
	\begin{abstract}
		Understanding the atomistic mechanism in graphene growth is crucial for controlling the number of layers or domain sizes to meet practical needs. In this work, focusing on the growth of graphene by chemical vapor deposition on copper substrates, the surface kinetics in the growth are systematically investigated by first-principles calculations. The phase diagram, predicting whether the growth mode is monolayer graphene or bilayer graphene under various experimental conditions, is constructed based on classical nucleation theory. Our phase diagram well illustrates the effect of high hydrogen pressure on bilayer graphene growth and clarifies the mechanism of the most widely used experimental growth approaches. The phase diagram can provide guidance and predictions for experiments and inspires the study of other two-dimensional materials with graphene-like growth mechanisms.
	\end{abstract}
	
	\maketitle
	
	\section{\Rmnum{1}. Introduction}
	\label{Intro}
	Graphene has sparked a wave of interest as a zero-bandgap two-dimensional material since its first successful exfoliation\cite{Novoselov-2004}. Exceptional electronic properties and impressive stability make it one of the most remarkable materials in electronic and energy-storage applications\cite{Morozov-2008, Lee-2008, Briggs-2010}. The properties of graphene strongly depends on the number of layers and the stacking order\cite{Yao-2022, Cai-2021}. Graphene monolayer exhibits excellent charge carrier mobility and optical transmittance, making it a promising candidate for applications in optoelectronic memory devices, electrodes and solar cells\cite{Bolotin-2008,Nair-2008, Bae-2010, Wang-2014, Xia-2009, Yu-2015}. Stacking of two or more graphene monolayers may lead to the emergence of novel physical properties, including tunable transport and optical properties\cite{Zhangyuanbo-2009,Mak-2009,Wangfeng-2008,YinJianbo-2016}, as well as the Mott insulating state\cite{Cao-2018-in} and superconductivity\cite{Cao-2018-sc}. Thus it is crucial to control the number of layers and the stacking order in graphene growth for large-scale practical applications. Understanding the growth mechanism of graphene with different layers is essential for achieving controllable graphene growth.
	
	Diverse growth methods have been developed to achieve the controllable growth of graphene\cite{Hernandez-2008, Shih-2011, Nyakiti-2012,Virojanadara-2008,Liu-2012}. Due to the ease of operation and a wide window of growth parameters, chemical vapor deposition (CVD) has an advantage over the complex and low-productivity exfoliation-stacking method\cite{Shih-2011,Hernandez-2008}, as well as the costly SiC sublimation method\cite{Virojanadara-2008, Nyakiti-2012}. In CVD growth of graphene, methane gas commonly serves as the carbon source with transition metals acting as the substrate\cite{Seah-2014}. Among the transition metals, copper is preferable due to its low carbon solubility making layer control easier\cite{Lopez-2004, Lixuesong-2009-nanoletter,Lixuesong-2009-science, Edwards-2012}. The deposited carbon source gas first decomposes into active carbon precursors, catalyzed by the metal substrate. Then the nucleation and growth proceed, the mechanism of which depends on the kinetic processes of the active carbon precursors.	
	
	\begin{figure}[b]
		\centering
		\includegraphics[width=\linewidth]{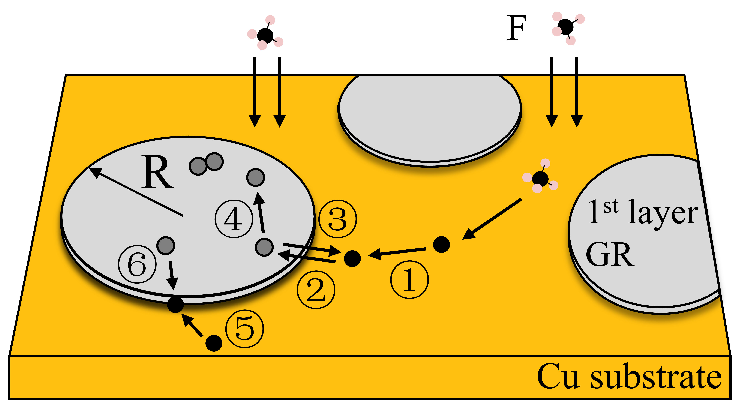}
		\caption{Schematic diagram of important kinetic processes involved in the BLG growth on a Cu substrate. The yellow platform denotes the Cu substrate. The gray circular disks represent the first-layer graphene islands, while the black and dark gray spheres represent carbon atoms adsorbed on the Cu substrate and at the graphene-Cu interface, respectively. }
		\label{F1}
	\end{figure}
	
	Normally, the conventional thin film growth can be described as the $wedding$-$cake$ mode, in which the nucleation of new layers occurs on top of the lower one\cite{Li2009,Ding2003}. However, the $inverse$ $wedding$-$cake$ (IWC) growth mode has been revealed in low-pressure CVD experiments during the bi-layer graphene (BLG) growth process\cite{Nieshu-2012, Fangwenjing-2013, Liqiongyu-2013, Chanchiachun-2018}. In this growth mode, the second layer graphene nucleates beneath and grows in competition with the first layer. The growth of the first layer would lead to a decrease in catalytic surface area and a shortage of carbon precursors, thereby inhibiting the growth of the second layer, which is known as the self-limiting growth. Moreover, the second layer would cease to grow due to the shortage of carbon precursors once the substrate is fully covered by the first layer.
	
	The unique graphene growth mechanism originates from the competing surface kinetic processes. Fig. \ref{F1} illustrates the surface kinetic processes involved in BLG growth, assuming that the growth unit is solely the C monomer. These processes can be divided into two categories. The first category includes the surface diffusion processes, namely on the clean substrate ($\#$1), across the step edge of the first layer graphene to the graphene-substrate interface ($\#$2) and the inverse one ($\#$3), and at the graphene-substrate interface ($\#$4). The second category involves the attachment processes that promote the growth of the first layer. They are labeled as $\#$5 and $\#$6, denoting the attachment to the first layer graphene from the uncovered region of the substrate and the graphene-Cu interface, respectively. 
	
	Many studies on the surface kinetics of graphene growth have focused on processes $\#$1 and $\#$5, providing insights into single layer graphene (SLG) growth\cite{Wuping-2010, Lipai-2017, Gaojunfeng-2011, Gaojunfeng-2011-2, Kim-2012, Chenwei-2015, Chenhua-2010, Riikonen-2012, Wuping-2015, Andersen-2019}. Beyond SLG growth, Zhang et al. conducted extensive calculations on the possible kinetic processes near the boundary of graphene island, and reported a much lower barrier of process $\#$2 at the H-terminated graphene edge, which is favorable for BLG growth\cite{Zhangxiuyun-2014}. Wei et al. also investigated the underlying atomistic mechanisms for BLG growth. They identified the critical island size required for the nucleation of the second layer and re-emphasized the effect of hydrogen gas on BLG growth\cite{Chenwei-2015}. However, a generic growth model that can serve as a guide for experiments under various conditions has not yet been developed. 
	
	In this paper, the surface kinetic properties of C monomers on Cu(111) surface are systematically investigated using first-principles calculations.
	Based on the full understanding of the surface kinetics, a phase diagram of graphene growth modes is constructed according to the classical nucleation theory. Using the phase diagram, the conditions for BLG growth are determined and the growth mechanisms of commonly used growth techniques are elucidated. This work would motivate further investigations into the IWC growth mechanism of other 2D materials.
	
	\section{\Rmnum{2}. Models $\&$ Methods}
	\label{method}
	
	The calculations are based on density functional theory (DFT) in the Perdew-Burke-Ernzerhof generalized gradient approximation (GGA-PBE) using the Vienna Ab initio Simulation Package (VASP), with Projector-Augmented Wave (PAW) methods\cite{Kresse-1996,Kresse-1996-2,Perdew-1996,Blochl-1994}. The semiempirical approach DFT-D2 is used to include the van der Waals (vdW) interactions\cite{Bucko-2010}. A cutoff energy of 650 eV for the plane wave basis set is employed for all calculations. The Cu(111) substrate is constructed based on the bulk Cu with an optimized lattice parameter of 3.634 \AA. Three slab models (M1-M3) are used for different aims in the simulation. The bare Cu(111) surface is modelled by a (5$\times$5) supercell consisting of a 5-layer slab and a 15 \AA \  vacuum layer (M1). The Cu(111) surface partially covered by graphene is modelled by a Cu substrate of (9$\times$4) supercell plus a (3$\times$4) graphene supercell on top of the Cu substrate (M2). The Cu(111) surface fully covered by graphene is modelled by introducing an additional (5$\times$5) graphene monolayer on the Cu substrate of (5$\times$5) supercell (M3). The graphene is streched slightly to match the lattice parameter of the Cu substrate in M2 and M3. 
	
	The in-plane Brillon zones of the (5$\times$5) supercell(M1/M3) and the (9$\times$4) supercell(M2) are sampled by a $\Gamma$-centered 3$\times$3 mesh and a $\Gamma$-centered 2$\times$4 mesh, respectively\cite{Methfessel-1989}. The bottom two layers of Cu slabs are fixed to mimic the bulk in all three models. The relaxations are carried out until forces on the free atoms are converged to 0.01 eV/\AA\ for structural optimizations. The energy barriers of the kinetic processes are calculated by the climbing image nudged elastic band (CI-NEB) method with the forces on the free atoms converged to 0.02eV/\AA \cite{Henkelman-2000}. These settings ensure a convergence in total energy of 0.1 meV/atom. 
	
	\section{\Rmnum{3}. Surface kinetics}
	\label{surfacekinetics}
	
	The main surface kinetic processes involved in the growth of BLG on Cu(111) substrates are shown in Fig. \ref{F1}. It is widely accepted that the C monomer is more stable on the subsurface of the Cu(111) substrate than on the surface\cite{Zhangxiuyun-2014,Chenwei-2015,Chenhua-2010,Riikonen-2012,Wuping-2015,Andersen-2019}. Using model M1, our calculations indicate that the former is energetically favored by about 0.5eV. Therefore, we denote the adsorption site on the subsurface of Cu(111) as $S_A$ and set it as a reference for adsorption energies at other sites. Based on the calculations using model M2, the adsorption of a C monomer near the edge of graphene islands becomes energetically more favorable. The energy gain is about 0.14 eV ($S_B$) or 0.01 eV ($S_C$), depending on whether the adsorption site is outside or inside the graphene covered zone. In contrast, when the adsorption site is located beneath the graphene islands but far from the edge ($S_D$), the calculation using model M3 shows that it becomes energetically less favorable by 0.02 eV.
	
	\begin{figure}[b]
		\centering
		\includegraphics[width=\linewidth]{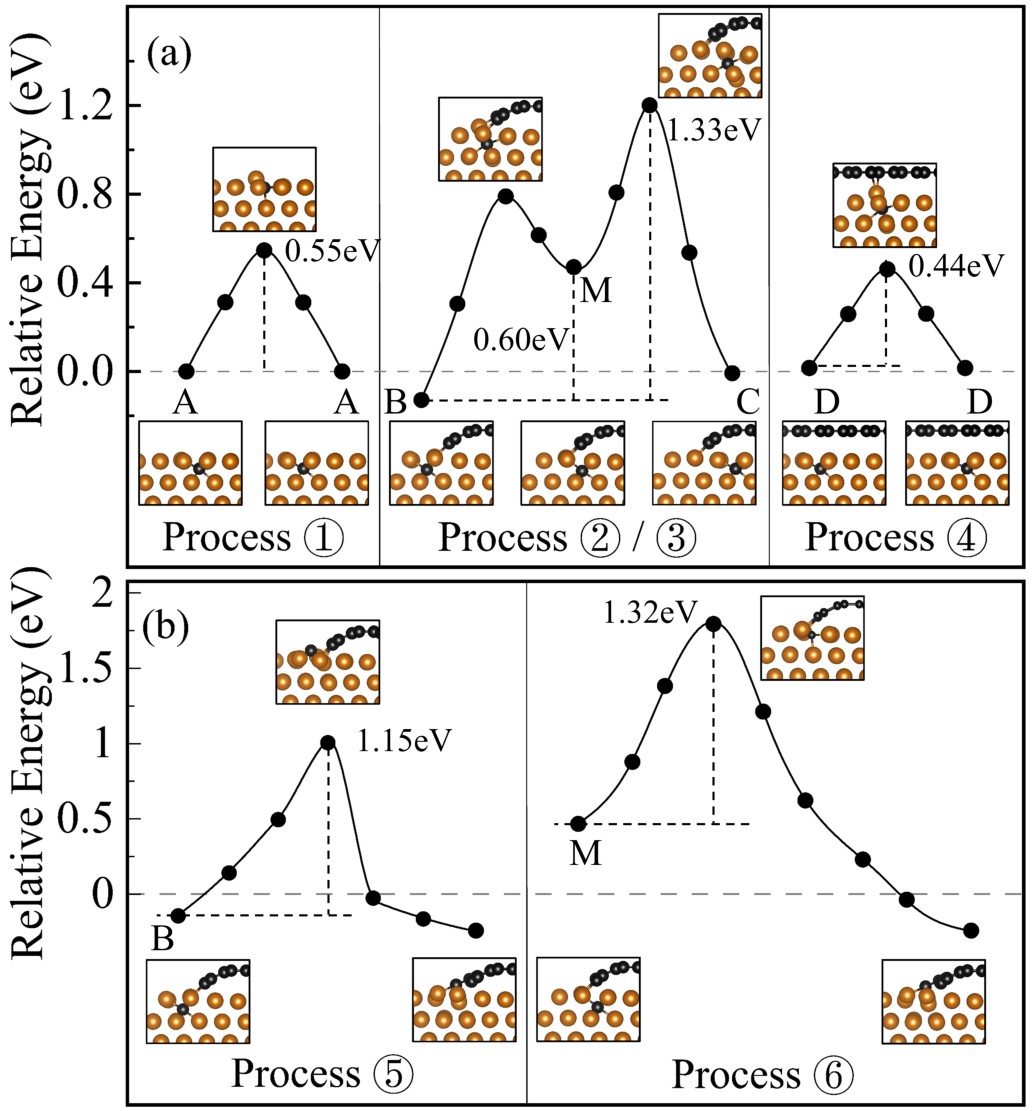}
		\caption{Energy profiles and optimized geometries involved in the C monomer (a) diffusion and (b) attachment processes in BLG growth on Cu(111) substrate, as indicated in Fig. \ref{F1}. The adsorption energy of the C monomer at $S_A$ is set as the reference. The process $\#$3 is the reverse of the process $\#$2. The configurations of the stable/metastable states and the transition states are given by the inset figures below and above  the energy curve, respectively.}
		\label{F2}
	\end{figure}
	
	The difference in adsorption energy at different sites ($S_A-S_D$) mainly arises from the substrate deformation induced by adsorption. The adsorption of a C monomer on the subsurface of a Cu(111) substrate leads to a local expansion near the adsorption site. At the proximity of the graphene islands, the interaction between the graphene edges and the substrate slightly lifts the surface Cu atoms. Therefore, the substrate exhibits less deformation after the adsorption of C monomer at $S_B$ and $S_C$ than at $S_A$, resulting in a lower adsorption energy. The adsorption of a C monomer at site $S_D$ costs more energy than at $S_A$ because the fully covered graphene inhibits the relaxation of the surface Cu atoms.
	
	We then focus on the four diffusion processes of C monomers. As shown in Fig. \ref{F2}(a), C monomers tend to diffuse on the subsurface of the Cu substrate, regardless of whether it is covered by graphene or not. The diffusion barriers on the uncovered and graphene fully covered Cu substrate are $\Delta G_1=0.55$ eV and $\Delta G_4=0.44$ eV, respectively. In comparison, the diffusion across the edge of the graphene island has a relatively high barrier. The energy barrier from the uncovered substrate to the graphene-covered substrate is $\Delta G_2=1.33$ eV, and the barrier of the reverse process is $\Delta G_3=1.20$ eV. The higher barrier across the edge can be attributed to the interaction between the dangling bonds of the graphene edge and the metal substrate. This interaction is also manifested by the bending of the graphene edge towards the substrate, as shown in the inset geometry configurations of Fig. \ref{F2}(a). 
	
	Next, we investigate the attachment processes on the substrate. As shown in Fig. \ref{F2}(b), the C monomer needs to overcome a barrier of $\Delta G_5=1.15$ eV to attach to the graphene edge from the uncovered substrate side, slightly less than the edge diffusion barrier $\Delta G_2$. For process $\#$6, the C monomer preferentially diffuses outward from beneath the graphene islands and then attaches to the edge. For comparison, we still obtain a direct attachment barrier of $\Delta G_6=1.92$ eV by imposing restrictions on the pathways of the C monomer. 
	
\begin{figure}[t]
		\centering
		\includegraphics[width=0.7\linewidth]{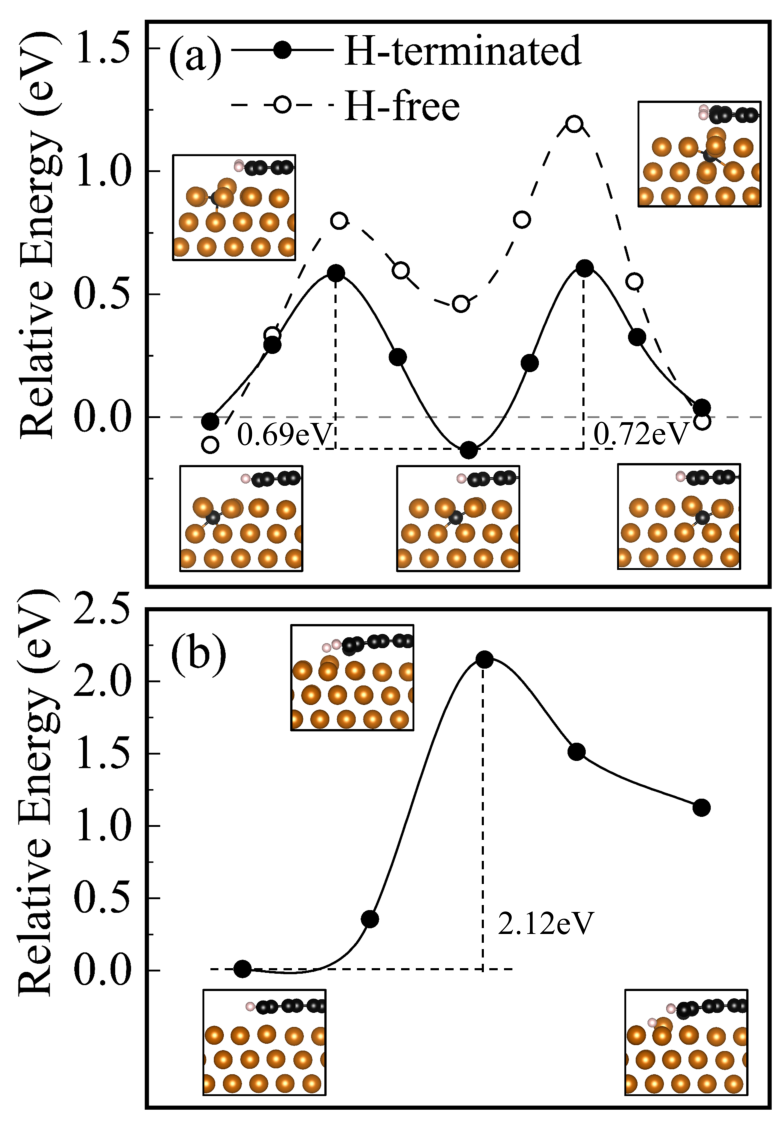}
		\caption{(a) Energy profiles and optimized geometries involved in the diffusion process that C monomer across the boundary of H-free (dashed line) and H-terminated (solid line) first-layer graphene islands. (b) Energy profile and optimized geometries involved in the H atom detachment process. The configurations of the stable/metastable states and the transition states are given by the inset figures below and above the energy curve, respectively.}
		\label{F3}
	\end{figure}

	We now examine the impact of hydrogen gas on the surface kinetics during graphene growth. H atoms are highly involved in graphene growth as hydrogen gas is comparatively easier to decompose than methane\cite{Zhangxiuyun-2014}. H atoms can saturate the edges of the first-layer graphene islands, reducing the interaction between the graphene edge and the substrate. This change facilitates the diffusion of C monomers across the island edge. As shown in Fig. \ref{F3}(a), the H termination of edges reduces the diffusion barriers $\Delta G_2$ and $\Delta G_3$ to 0.72 eV and 0.69 eV, respectively. Meanwhile, the diffusion barriers of $\Delta G_1$ and $\Delta G_4$ remain nearly unaffected by the H termination of edges. On the other hand, the detachment of H atoms is a prerequisite condition for the attachment processes at the H-terminated edge. Fig. \ref{F3}(b) shows that the barrier for H atom detachment from the H-terminated edge is 2.12 eV, which is higher than the attachment barriers $\Delta G_5$ and $\Delta G_6$. Therefore, the detachment of H atom from edges becomes the rate-limiting step for the growth of the first-layer graphene islands. It is much easier for C monomers to accumulate beneath the first-layer islands under high hydrogen pressure due to the difficulty of attachments, which increases the probability of second layer nucleation. 
	
	There have been many studies on the simulation of the kinetic properties of C monomers. The value of $\Delta G_1$ in the current work is in agreement with the reported barriers of C monomer diffusion on Cu substrate\cite{Wuping-2010, Chenhua-2010, Chenwei-2015, Riikonen-2012, Wuping-2015, Lipai-2017}. The value of $\Delta G_4$ in the current work is higher than the reported results, which are approximately 0.3 eV.\cite{Zhangxiuyun-2014, Chenwei-2015, Lipai-2017}. This difference possibly results from the larger (5$\times$5) supercell used in our calculation. Besides, the previous work has investigated the processes that occur at the edges of islands, as well as the effect of H-terminated edges on these barriers. The reported hopping barriers of processes near the edge are slightly smaller than our results\cite{Zhangxiuyun-2014, Lipai-2017}. This difference may originate from the tensile strain in the graphene introduced by the Cu substrate in our calculations, as we have demonstrated that a larger lattice of the system leads to a higher barrier for the same kinetic path.
	
	\section{\Rmnum{4}. Phase diagram}
	\subsection{A. Concentrations of C monomers}
	\label{4A}
	
	With a complete understanding of the surface kinetic processes involved in growth, we are able to achieve deeper insights into the graphene growth mode using rate equation approaches. Two approximations are used in our model. Firstly, we ignore the details of gas decomposition and assume the C monomers arrive at the substrate at a fixed deposition rate of $F$ per $A_0$ in the normal direction, where $A_0$ is the area of an adsorption site on the substrate. It is noteworthy that the deposition rate on the region covered by C clusters is zero, as the gas decomposition requires the aid of the catalytic substrate in experiments. Secondly, we assume the growth unit is the C monomer and its mean residence time before desorption is long enough to incorporate into clusters. As shown in Fig. \ref{F1}, we start with a model where circular first-layer graphene islands have formed on the Cu substrate. The average radius of the first-layer islands and half of the average island-island separation are denoted as $R$ and $L$, respectively, both in units of $\sqrt{A_0}$. 
	
	The deposited C monomers can diffuse across or attach to the boundary of the first-layer graphene islands. The corresponding rate of the six processes is given by $\nu_j=\nu_0{\exp (-\Delta G_j/kT)}$, where $\Delta G_j$ is the kinetic barrier of process $j$ $(j=1-6)$, $\nu_0$ is the attempt frequency, $k$ is the Boltzmann constant and $T$ is the experimental temperature. It is reasonable to assume that the growth is limited by the attachment, since the kinetics of diffusion processes are much more rapid. In this case, the steady-state C monomer concentration is easily reached.
	
	According to the rate equation, the steady-state concentration of C monomer beneath the first-layer graphene islands remains constant due to the absence of direct atomic deposition at the graphene-substrate interface. We denote this concentration as $\theta_1$, which also represents the probability of an adsorption site beneath the first-layer graphene islands being occupied by a C monomer. On the contrary, the steady-state concentration of C monomer outside the graphene islands is position-dependent, influenced by the deposition rate and diffusion kinetics. Here we are only interested in the C monomer concentration at the edge outside the graphene islands, which is denoted as $\theta_1^{\prime}$. 
	
	The total number of C monomers beneath the first-layer graphene islands $N_1$ is only affected by the kinetic processes $\#$2, $\#$3 and $\#$6. Therefore, it evolves as follows,
	\begin{equation}
		\frac{\partial N_1}{\partial t}\bigg|_{R}=2\pi R\theta_1^{\prime}\nu_2-2\pi R\theta_1\nu_3-2\pi R\theta_1\nu_6.  \label{eq1}
	\end{equation}
	The steady-state value of $N_1$ at a certain first-layer island radius $R$ remains constant over time, which gives
	\begin{equation}
		\theta_1=\frac{\nu_2}{\nu_3+\nu_6}\theta_1^{\prime}.  \label{eq2}
	\end{equation}
	
	Similarly, the steady-state value of the total number of C monomers on the exposed Cu substrate is also constant if only the first-layer radius $R$ is fixed. Therefore, the deposited atoms contribute solely to the growth of first-layer islands at steady state and fixed $R$, which follows that
	\begin{equation}
		\frac{d\left(\pi R^2\right)}{dt}=\pi\left(L^2-R^2\right)F. \label{eq3}
	\end{equation}
	On the other hand, the first-layer islands expand in size through the attachment processes $\#$5 and $\#$6 according to the boundary condition. It gives that
	\begin{equation}
		\frac{d\left(\pi R^2\right)}{dt}=2\pi R\left(\theta_1^{\prime}\nu_5+\theta_1\nu_6\right).  \label{eq4}
	\end{equation}
	
	Combining Eq. (\ref{eq2}) - (\ref{eq4}), we have
	\begin{equation}
		\theta_1=\frac{L^2-R^2}{2R\alpha\Gamma},  \label{eq5}
	\end{equation}
	and the number of C monomers beneath the first-layer graphene islands at steady state is
	\begin{equation}
		N_1=\pi R^2\theta_1=\frac{\pi R\left(L^2-R^2\right)}{2\alpha\Gamma}, \label{eq6}
	\end{equation} 
	where 
	\begin{equation}
		\alpha=\frac{\nu_3\nu_5+\nu_2\nu_6+\nu_5\nu_6}{\nu_1\nu_2},\  \Gamma=\frac{\nu_1}{F}. \label{eq7}
	\end{equation}
	As shown in Eq. (\ref{eq7}), $\alpha$ is merely determined by the surface kinetics while $\Gamma$ depends on both the surface kinetics and the deposition rate. In this way, the complicated surface kinetic properties and growth conditions are described by the two dimentionless factors $\alpha$ and $\Gamma$. Such reduction makes the growth phase diagram possible. Specifically, by using the kinetic results in Sec. \Rmnum{3}, we have $\alpha_{\rm H-free}=1.5\times 10^{-2}$ and $\alpha_{\rm H-rich}=1.9\times 10^{-6}$ for the experiments under low and high $H_2$ partial pressure at $T=1300$ K, respectively. 
	
	\begin{table*}[t]
		\begin{center}
			\caption[content...]{Summary table of the average binding energy per atom in the first-layer critical nucleus ($E^{\prime}_{i}$) and second-layer critical nucleus($E_{i}$), the binding energy difference ($\Delta E_{i}$), the coefficients ($b_{i}$ in Eq. (\ref{eq9}) and $C_{i}$ in Eq.(\ref{eq11})) and the function $f_{i}(x)$ for different critical nucleus size $i$. The data of $E_{i}$ and $E^{\prime}_{i}$ are reproduced from Ref. \cite{ZhongLixiang-2016} and Ref. \cite{Wuping-2014}, respectively. A higher value indicates a more stable cluster binding. The results of $b_{i}$ and $C_{i}$ are gained by setting $T=1300$ K, $\sigma_{i}=2$, $\sigma_s=10$, $\Theta_C=0.2$ and $R_n=3$. } 
			\label{T1}
			\renewcommand{\arraystretch}{1.4}
			\begin{tabular}{m{1cm}<{\centering}m{1.5cm}<{\centering}m{1.5cm}<{\centering}m{1.5cm}<{\centering}m{1.5cm}<{\centering}m{1.5cm}<{\centering}m{8.3cm}<{\centering}}
				\hline\hline
				$i$ & ${E_{i}}^{\prime}$(eV) & $E_{i}$(eV) & $\Delta E_{i}$(eV) & $b_{i}$ & $C_{i}$ & $f_{i}(x)$ \\	\hline
				1 & 0 & 0 & 0 & 1.18 & 22.54 & $\frac{1}{4}x^6-\frac{1}{2}x^4+\frac{1}{4}x^2$\\	\hline
				2 & 1.20 & 0.43 & -0.77 & 0.09 & 3.15$\times10^{-5}$ & $-\frac{1}{5}x^8+\frac{2}{3}x^6-x^4+\frac{8	}{15}x^3$\\	\hline
				3 & 1.11 & 0.29 & -0.82 & 0.07 & 8.94$\times10^{-9}$ & $\frac{1}{6}x^{10}-\frac{3}{4}x^8+\frac{3}{2}x^6-x^4\ln x-\frac{11}{12}x^4$\\	\hline
				4 & 1.31 & 0.63 & -0.68 & 0.03 & 1.55$\times10^{-9}$ & $-\frac{1}{7}x^{12}+\frac{4}{5}x^{10}-2x^8+4x^6-\frac{128}{35}x^5+x^4$\\	\hline
				5 & 1.33 & 0.70 & -0.63 & 0.02 & 4.35$\times10^{-11}$ & $\frac{1}{8}x^{14}-\frac{5}{6}x^{12}+\frac{5}{2}x^{10}-5x^8+5x^6\ln x+\frac{65}{24}x^6+\frac{1}{2}x^4$\\	\hline
				6 & 1.38 & 0.77 & -0.61 & 0.01 & 4.76$\times10^{-13}$ & $-\frac{1}{9}x^{16}+\frac{6}{7}x^{14}-3x^{12}+\frac{20}{3}x^{10}-15x^8+\frac{1024}{63}x^7-6x^6+\frac{1}{3}x^4$  \\	\hline\hline
			\end{tabular}
		\end{center}
	\end{table*}
	
	\subsection{B. Monomers beneath islands}
	\label{4B} 
	The nucleation of second-layer islands is more favorable if $N_1$ is larger than the size of a critical nucleus $i$. Therefore, a rough estimation of $N_1$ is necessary in order for preliminary screening of the conditions for BLG growth.
	
	According to Eq. (\ref{eq6}), $N_1$ has a maximum value of $N_1^{max}=\sqrt{3}\pi L^3/9\alpha\Gamma$ at $R=\sqrt{3}L/3$. The half of the island-island separation $L$ is correlated with the concentration of first-layer nuclei $\theta_s$ by $\pi L^2\theta_s=1$.	In some situations the nucleation density is controlled by seeds or specific templates\cite{Haoyufeng-2013}, making the size of $L$ case-by-case. Here we assume the more general spontaneous nucleation. In this case, the first-layer nuclei have a saturation concentration $\theta_s\propto\Gamma^{-i/(i+2)}$ and thus $L$ depends on $\Gamma$ as follows\cite{Venables1973, Venables1984},
	\begin{equation}
		L=b_{i}\Gamma^{\frac{i}{2i+4}}, \label{eq8}
	\end{equation}
	where $b_{i}$ is a proportional coefficient. Therefore, we obtain the maximum value of the total C monomers beneath the first-layer island as follows, 
	\begin{equation}
		N_1^{max}=\frac{\sqrt{3}\pi b_i^{\frac{2i+4}{i}}}{9\alpha L^{\frac{4-i}{i}}}.  \label{eq9}
	\end{equation}
	
	According to the classical nucleation theory (See Supplementary Material), the expression of $b_{i}$ is as follows,
	\begin{equation}
		b_{i}=\pi^{-\frac{1}{2}}[\frac{\exp (-i{E_{i}}^{\prime}/kT)\sigma_s^{i+1} }{(i+2)\sigma_{i}\Theta_C}]^{\frac{1}{2i+4}}, \label{eq10}
	\end{equation}
	where ${E_{i}}^{\prime}$ is the average binding energy per atom of the first-layer critical nucleus, $\Theta_C$ is the coverage when the nucleus density is saturated, $\sigma_{i}$ and $\sigma_s$ are the average number of attachment sites of a critical nucleus and stable nucleus, respectively. By setting $T=1300$ K, $\sigma_{i}=2$, $\sigma_s=10$, $\Theta_C=0.2$, and adapting the binding energies reported by Zhong et al.\cite{ZhongLixiang-2016}, the values of $b_{i}$ for $i=1-6$ are listed in Table \ref{T1}. 
	
	Combining Eq.(\ref{eq8}) - Eq.(\ref{eq10}), we can determine the range of $L$ that satisfies $N^{max}_1>i$. For $i<4$, it requires that $L$ is smaller than $10^4$. For $i=4$, $N^{max}_1$ is always smaller than $i$. For $i>4$, it requires that $L$ is larger than $10^{19}$ or 10 for H-free or H-rich situations, respectively. In the typical experimental nucleation density range of 10$^{-2} - 10^4$ mm$^{-2}$ (See Supplementary Material), $L$ is approximately within $10^4-10^7$\cite{Zhaopei-2016,Chanchiachun-2018}. Thus BLG growth is possible when first-layer graphene islands are terminated by H atoms for a critical nucleus size $i>4$. 
	
	\subsection{C. Second Layer Nucleation}
	\label{4C}
	We now consider the second-layer nucleation beneath the first-layer graphene islands, assuming the critical size of the second-layer nucleus is still $i$, the same as the first-layer nucleus. According to the classical nucleation theory, the nucleation rate per unit area can be written as $J=\sigma_{i}\nu_4\theta_1\theta_{i}$, where $\theta_{i}$ is the concentration of clusters of size $i$ at steady state\cite{Venables1973, Venables1984}. The expression of $\theta_{i}$ is given by $\theta_{i}= \theta_1^{i}\exp(iE_{i}/kT)$, where $E_{i}$ is the average binding energy per atom of the second-layer critical nucleus\cite{Walton1962}. 
	
	\begin{figure}[b]
		\centering
		\includegraphics[width=0.8\linewidth]{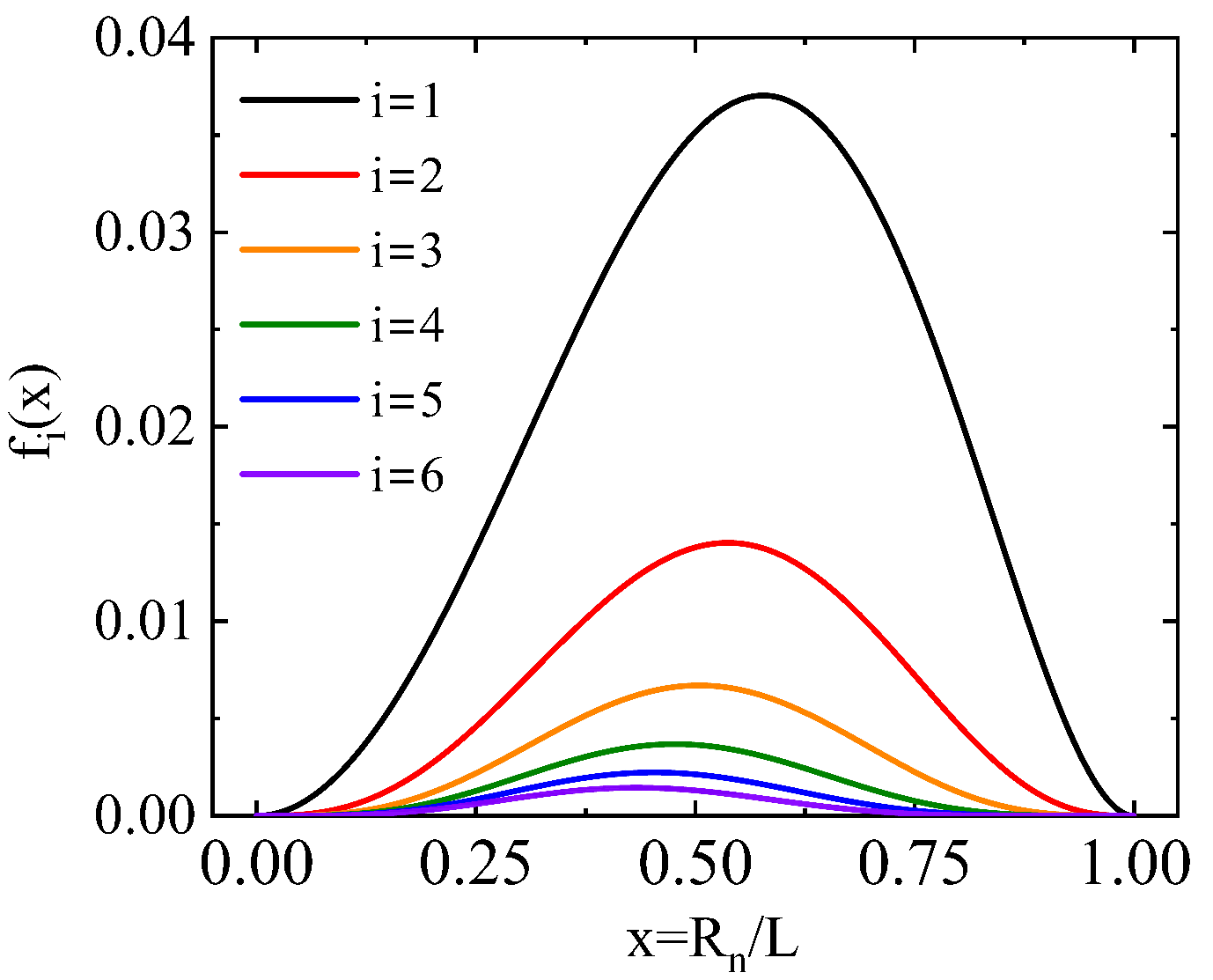}
		\caption{Variation of $f_{i}(x)$ with $x$ for different values of $i$.}
		\label{F4}
	\end{figure}
	
	Obviously, the number of second-layer nuclei increases as the first-layer islands grow, until $R$ equals to $L$ and the first-layer islands coalesce. Subsequently, the growth unit is no longer supplied due to the absence of exposed catalytic substrate. The upper limit of the nucleus number beneath the first-layer graphene islands can thus be obtained as $I_{i}(L)=\int_{R_n}^{L}J\pi R^2\frac{dt}{dR} dR$, where $R_n$ represents the minimal radius of the well-defined first-layer graphene island from which the second-layer nucleation begins to form. For a nucleus with a radius $R_n$, its size $N_n\simeq3\sqrt{3}R_n^2/2$. Here we choose the minimal first-layer island involved in the second-layer nucleation as the C$_{24}$ cluster containing seven 6-membered rings, which means $R_n=3$. Combining with Eq. (\ref{eq3}), we have 
	\begin{equation}
		I_{i}=\int_{R_n}^{L}J\pi R^2\frac{dt}{dR} dR=\frac{C_{i}f_{i}(x)}{(\alpha R_n)^{i+1}}, \label{eq11}
	\end{equation}
	where $x=R_n/L$ and the proportional coefficient $C_{i}=\frac{2\lambda}{(i+2)\Theta_C}(\frac{\sigma_s}{2\pi})^{i+1}\exp(i\Delta E_{i}/kT)$. Here $\lambda=\nu_4/\nu_1$, and $\Delta E_{i}=E_{i}-{E_{i}}^{\prime}$ denotes the difference in the average binding energies per atom of the critical nuclei at the graphene-substrate interface and on the Cu substrate. A negative $\Delta E_{i}$ indicates that the cluster is more favorable to form on the bare substrate than beneath the first-layer graphene islands. By using the forementioned values and setting $\lambda=2.67$ based on the kinetic properties obtained in Sect. \Rmnum{3}, the values of $C_{i}$ for $i=1-6$ are listed in Table \ref{T1}. The expressions of $f_{i}(x)$ for $i=1-6$ are also listed in Table \ref{T1}. Fig. \ref{F4} shows the variation of $f_{i}(x)$ with $x$. It can be identified that $f_{i}(x)$ reaches its maximum at a moderate value of $x$. Therefore, the number of second-layer nuclei has a maximum at a moderate nuclei concentration. As for the effects of the parameters $\alpha$ and $\Gamma$ on $I_i$, it is noteworthy that $I_i$ is inversely proportional to the dimensionless factor $\alpha$, while the influence of $\Gamma$ on $I_i$ is mediated by $x$ and $f_i(x)$.

	\subsection{D. Growth Modes}
	\label{4D}
	The growth mode can now be estimated according to the value of $I_{i}$. If $I_{i} \textless 1$, there is little possibility that a second-layer nucleus forms before the coalescence of first-layer islands. Only if $I_{i} \textgreater 1$, the second-layer nucleation will occur and the BLG growth becomes possible. Therefore the condition $I_{i}=1$ gives the border between SLG growth and BLG growth. The phase diagram of the growth mode can thus be constructed using the values in Table \ref{T1}. 
	
	\begin{figure}[b]
		\centering
		\includegraphics[width=\linewidth]{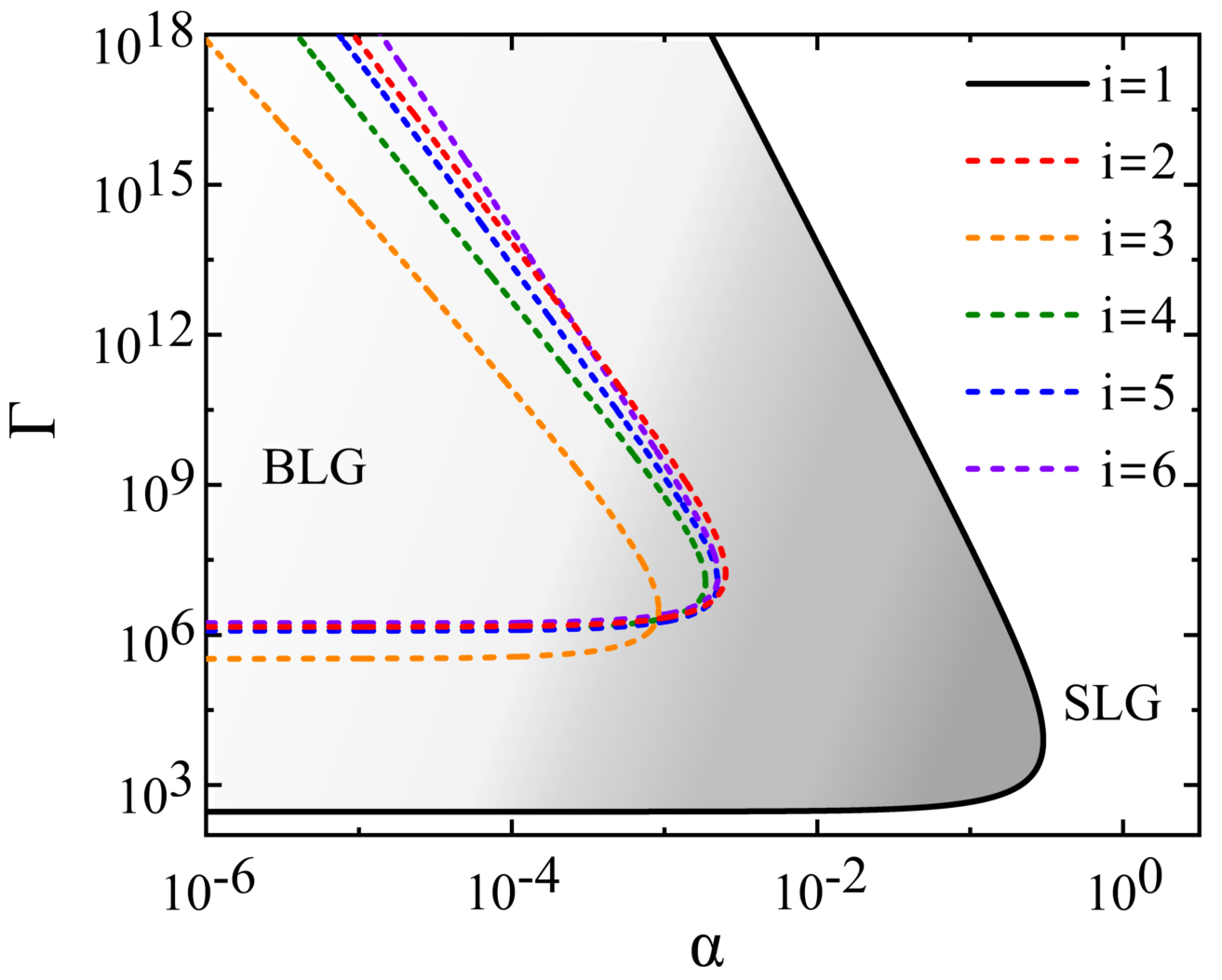}
		\caption{Growth phase diagram of graphene growth on the Cu substrate constructed using $\alpha$ and $\Gamma$. The gradient-filled area and colorless area represent BLG growth and SLG growth, respectively. The boundary curves separating the BLG region and the SLG region are determined by comparing $I_i$ and 1. The boundary curves of $i=2-6$ are depicted based on the values given in Table. \Ref{T1}} 
		\label{F5}
	\end{figure}
	
	We first consider the simplest situation of $i=1$ as the binding energy is zero. The black solid line in Fig. \ref{F5} represents the border curve of BLG and SLG growth when $i = 1$, which divides the diagram region into two sections.
	The gradient-filled area in the top left corner of the diagram is suitable for BLG growth, while the remaining colorless area only permits SLG growth. The diagram is very similar to the one that predicts the layer-by-layer growth and island growth we reported before\cite{Shudajun-2017}, indicating that the IWC growth is analogous to the interfacial growth.
	
	Now it is easy to identify the influence of the reduced parameter $\alpha$ on the growth mode. As shown in Fig. \ref{F5}, only SLG growth occurs when $\alpha$ is larger than a critical value. When $\alpha$ is smaller than the critical value, BLG growth can occur within a moderate range of $\Gamma$ whereas SLG growth is reentrant when the value of $\Gamma$ is sufficiently small or large. Furthermore, the range of $\Gamma$ for BLG growth becomes larger for smaller $\alpha$, indicating that a larger $\alpha$ favors SLG growth while a smaller $\alpha$ favors BLG growth. According to Eq. (\ref{eq7}), a larger value of $\alpha$ implies more rapid attachment kinetics at the island edge or more sluggish diffusion into the graphene-covered area. Therefore, the concentration of C monomers beneath the first-layer graphene islands is always deficient to form the second-layer nucleus, resulting in the advantage of SLG over BLG. Conversely, a smaller $\alpha$ generally indicates a smaller attachment kinetic rates as well as a smaller outward diffusion rate than the intralayer diffusion rate, which favors the BLG growth. 
	
	The influence of $\Gamma$ on the growth mode is nonmonotonic. As discussed above, the SLG growth occurs only for a very small or large value of $\Gamma$ if $\alpha$ is not large enough. The growth of SLG at small $\Gamma$ corresponds to a very high density of first-layer nuclei, so that the second-layer nucleation hardly occurs before the coalescence of the first-layer islands. Conversely, the SLG growth at large $\Gamma$ is mediated by the low second-layer nucleation rate due to the low concentration of C monomers.
	
	The size of the critical nucleus varied with the growth condition also influences the growth mode. When $i>1$, the value of $\Delta E_i$ is no longer zero and the coefficients $b_i$ and $C_i$ become dependent on the temperature. The border curves for $i=2-6$ at $T=1300$ K are represented by the colored dashed curves in the phase diagram. The shrinkage of BLG growth area with larger $i$ originates mostly from the negative $\Delta E_{i}$ that results into smaller $C_i$ and smaller nucleation number $I_i$. There is little variations among the border curves for $i=2\sim6$, with an exception of $i=3$ because the magnitude of $\Delta E_{3}$ is largest than the others. It suggests an additional way to regulate the growth mode by tuning the binding energy of  nuclei through substrate engineering. 

	\section{\Rmnum{5}. Discussion}
	In the framework of the phase diagram, some strategies can be predicted to control the growth process. For instance, BLG growth can be intentionally selected based on three strategies. Firstly, a relatively low value of $\alpha$ is beneficial, as illustrated in the diagram. The value of $\alpha$ can be decreased by reducing the interaction between the edges of graphene islands and the substrate. For example, hydrogen termination of edges can reduce the value of $\alpha$ from $\alpha_{\rm H-free}=1.5\times 10^{-2}$ to $\alpha_{\rm H-rich}=1.9\times 10^{-6}$ based on our calculations. Secondly, a moderate value of $\Gamma$ is beneficial for BLG growth. Therefore, it is crucial to pay close attention to the gas flux and the deposition rate in experiments. Finally, enlarging the area of BLG growth in the diagram by changing the border is also a strategy for BLG growth. A possible approach is to increase $\Delta E_i$ through substrate modification, which would enhance the coefficient $C_i$ and thus the BLG growth area in the phase diagram.

	Up to now, many experiments have reported good yields of both SLG and BLG. Single crystal SLG has been successfully grown in experiments by dynamically controlling the gas ratio or pressure\cite{Lixuesong-2010, Chanshihhao-2013, Chaitoglou-2016,Sunluzhao-2016, Wutianru-2016, Zhouhailong-2013,Guowei-2016}, domain orientation control\cite{Chenwei-2012,Shi-2021}, or pretreating the substrate through coating\cite{LuoBirong-2017}, annealing\cite{Nguyen-2019,Luoda-2019}, polishing\cite{Yaowenqian-2022, Braeuninger-2016}, alloying\cite{Huangming-2018,Liuyifan-2018} or oxidizing\cite{SunLuzhao-2021-AcsNano,Zhouhailong-2013,Srinivasan-2018,Ganlin-2013, Eres-2014, Haoyufeng-2013}. The key points in these methods are suppressing nucleation density and improving growth rate while the self-limiting growth regime remains unchanged. These methods correspond to a large $\alpha$ and $\Gamma$, consistant with the diagram. 	
	Growth strategies for BLG are more diverse and complicated in experiments. Under a high H$_2$ partial pressure, the graphene edges were passivated by H atoms, which significantly reduce the value of $\alpha$ in the diagram based on our kinetic calculations. Both AB-stacking and twisted BLG are synthesized by dynamically adjusting the gas ratio in experiments\cite{Lim-2020, SunLuzhao-2021-NC,Zhaopei-2016}. Substrate engineering by alloying\cite{Yinghao-2019, Huangming-2020} or oxidation\cite{LiuBing-2019} to improve catalytic performance reduces the value of $\Gamma$ in the diagram, which is useful in low-pressure CVD experiments. The Cu pocket method\cite{Haoyufeng-2016, Lixuesong-2009-science, LiuBing-2019} and two-step growth\cite{Yankai-2011,LiuLixin-2012} provide additional catalytic surfaces that aid in the supply of growth units, which improves the adatom concentration beneath the first layer, yields the same results as reducing $\alpha$. The application of gas etching reduces the growth rate of first-layer islands\cite{Qizhikai-2018}, while the oxidation of substrate decreases the density of first-layer nucleation sites\cite{Haoyufeng-2016}. These strategies extend the time required for the full coverage of initial layer islands and expand the area available for the growth mode of BLG in the diagram while keeping $i$ constant.

	\begin{figure}[b]
		\centering
		\includegraphics[width=0.9\linewidth]{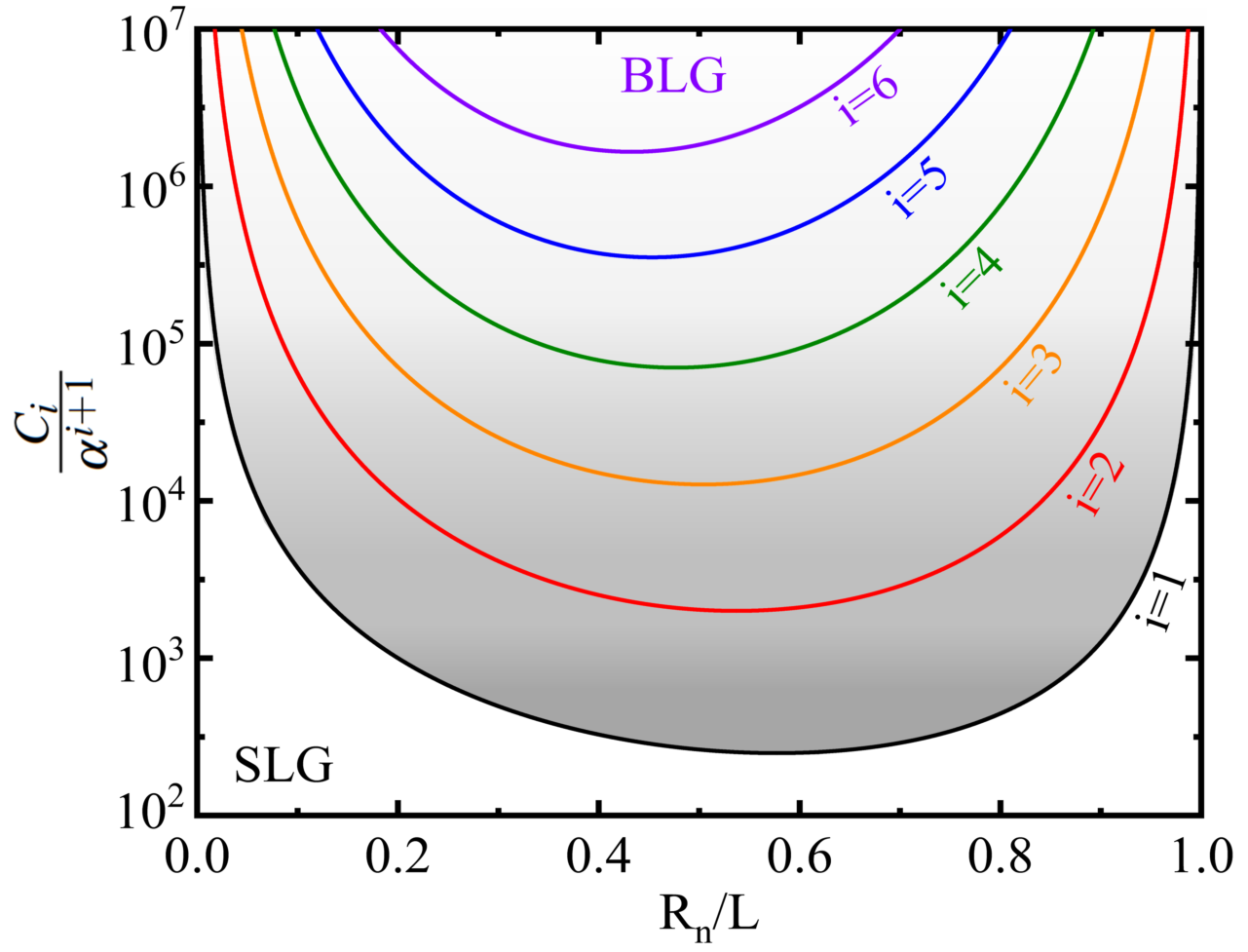}
		\caption{Growth phase diagram of graphene growth on the Cu substrate constructed using $C_i/\alpha^{i+1}$ and $R_n/L$. The gradient-filled area and colorless area represent BLG growth and SLG growth, respectively. The lines separating the BLG region and the SLG region are determined by comparing $I_i$ and 1.} 
		\label{F6}
	\end{figure}

	The above discussions are all based on Fig. \ref{F5} which is depicts the growth mode on the Cu(111) substrate. An alternative substrate would result in completely different surface kinetics and nucleation preference. Moreover, when the growth temperature varies, not only the values of $\alpha$ and $\Gamma$ change, the border curve also shifts. It is not straightforward to use the current phase diagram to cover these changes. A new diagram may be more helpful in the complicated situation. Fig. \ref{F6} shows the phase diagram constructed using $C_i/\alpha^{i+1}$ and $R_n/L$, which is effectively similar to the aforementioned one. Only SLG growth is possible when $C_i/\alpha^{i+1}$ is sufficiently small. When $C_i/\alpha^{i+1}$ is larger than the critical value, SLG growth occurs if $L$ is extremely large or small. Correspondingly, BLG growth occurs at a moderate value of $L$ as $C_i/\alpha^{i+1}$ increases. Beyond graphene, the phase diagram shown in Fig. \ref{F6} can also shed light on the growth control of other 2D materials with IWC growth mode mediated by the expression of $C_i$ and $L$. 

	\section{\Rmnum{6}. Conclusion}
	In summary, the mechanism of graphene growth on the Cu substrate is comprehensively studied by a complete analysis of the atomic kinetic processes and the construction of the growth mode diagram. Our diagram identifies the necessary conditions for BLG growth and gives the criteria that distinguish BLG growth from SLG growth. Furthermore, experimental strategies for SLG and BLG growth can be elucidated and summerized using the diagram. Our findings can serve as a guide for new designs of experiments and optimization of experiment parameters, but also inspire other 2D material growth studies with IWC growth mode.
	
	\begin{acknowledgments}
	\section{acknowledgments}
		The numerical calculations were carried out at the High Performance Computing Center of Nanjing University. This work was supported by the National Natural Science Foundation of China (Grants No. 12274211).
	\end{acknowledgments}

	\section{\Rmnum{7}. Appendix}
	\subsection{A. The saturation concentration of first-layer nuclei}
	To determine the saturation concentration of first-layer nuclei, we start by assuming nucleation is spontaneous, with only C monomers being mobile on the substrate. The two assumptions presented in the main text are also employed for simplicity. The C monomers arrive at the substrate due to deposition at a rate $F$ and diffuse on the substrate with a rate $\nu_1$. Adsorbed C monomers may re-desorb and the mean residence time for a monomer is given by $\tau=\frac{1}{\nu_0}\exp(E_a/kT)$, where $\nu_0$ is the vibrational frequency, $E_a$ is the adsorption energy of a C monomer, $k$ is the Boltzmann constant and $T$ is the experimental temperature. The remaining C monomers diffuse on the substrate until they encounter other monomers and form dimers. The dimers capture more monomers and grow step-by-step into larger clusters with size $n$. The dimers and clusters also have chance to decay. Clusters of size $n \textgreater i$ tend to grow rather than decay. Thus $i$ is referred to as the critical nucleus size. All clusters with size larger than $i$ are considered stable nuclei. 
	
	The concentration of C monomer on the bare substrate far away from any capture zone, $\theta_0$, increases due to deposition at the rate $F$ and decreases due to re-desorption at rate $1/\tau$, formation of C dimer at rate $U_1$ (each dimer removing two monomers), collection into unstable clusters at rate $U_n$ ($2\leq n \leq i$) and stable nuclei at rate $U_s$. Thus we have
	\begin{equation}
		\frac{d\theta_0}{dt}=F-\frac{\theta_0}{\tau}-2U_1-\sum\limits_{n=2}^{i} U_n- U_s. \label{eqA1}
	\end{equation}
	At the steady state $d\theta_0/dt=0$, thus the right-hand side of Eq. (\ref{eqA1}) is equal to zero. When the mean stay time $\tau$ is sufficiently long and the terms $U_1$ and $U_n$ are considerably smaller than the $U_s$ term, the steady value of $\theta_0$ is determined by the balance between the deposition rate and the capture rate of C monomers by the stable nuclei, $i.e.$, $F=U_s$. The capture rate of C monomers by the stable nuclei is given by $U_s=\sigma_s\nu_1\theta_0\theta_s$, where $\sigma_s$ is the average capture number of the stable nuclei and $\theta_s$ is the concentration of stable nuclei. The steady-state solution of Eq. (\ref{eqA1}) yields
	\begin{equation}
		\theta_0=\frac{F}{\sigma_s \nu_1\theta_s}. \label{eqA2}
	\end{equation}
	Note that $\theta_0$ given in Eq. (\ref{eqA2}) is the concentration of C monomers far away from the capture zones of stable nuclei. The concentration of C monomer within the capture zones of stable nuclei is much smaller than $\theta_0$. 
	
	The concentration of stable nuclei, $\theta_s$, increases at rate $U_i$ due to new nucleus formation and decreases at rate $U_c$ due to the coalescence of nuclei. The term $U_c$ usually ensures a upper limit for $\theta_s$. Here, we consider that the concentration of nuclei reaches saturation before the coalescence of nuclei occurs due to the self-limiting mechanism. In this case we neglect the term $U_c$ and focus on the term $U_i$ only. 
	
	Since the C monomer concentration in the near vicinity of the stable nuclei is reduced, nucleation exclusion zones that prohibit nucleation appear exist around stable nuclei and expand as the nuclei grow. The probability of a site being outside nucleation exclusion zones $P$ decreases with time exponentially, $i.e.$, $p=\exp[-f(t)]$, where $f(t)=Kt^3$ increases with time in a power law\cite{Markov-2003}. The increasing rate of the stable nuclei concentration is thus as follows
	
	\begin{equation}
		\frac{d\theta_s}{dt}=U_i=pJ=\sigma_i \nu_1 \theta_0\theta_i\exp[-f(t)].  \label{eqA3}
	\end{equation}
	The concentration of critical size nuclei $\theta_i$ is given by $\theta_i=\theta_0^i\exp(iE_i^{\prime}/kT)$\cite{Walton1962}, where $E_i^{\prime}$ is the average binding energy per atom in the first-layer critial nucleus. Combining Eq. (\ref{eqA2}) $, $ Eq. (\ref{eqA3}) turns into
	\begin{equation}
		\theta_s^{i+1}d\theta_s=\frac{\sigma_i}{\sigma_s^{i+1}}\Gamma^{-i}\exp(iE_i^{\prime}/kT)F\exp[-f(t)]dt. \label{eqA4}
	\end{equation}
	The saturation nucleation concentration is obtained by integrating $t$ from 0 to $\infty$ as follows
	\begin{equation}
		\theta_s=[(i+2)\frac{\sigma_i}{\sigma_s^{i+1}}\exp(iE_i^{\prime}/kT)\Theta_c]^\frac{1}{i+2}\Gamma^{-\frac{i}{i+2}}, \label{eqA5}
	\end{equation}
	where $\Gamma=\nu_1/F$ and $\Theta_c=\int F\exp[-f(t)]dt$ is the coverage of first-layer nuclei once its concentration reaches saturation. The half of the first-layer island-island separation $L$ is correlated with $\theta_s$ by relation $\pi L^2\theta_s=1$. Thus the expression of $L$ is as follows,
	\begin{equation}
		L=\pi^{-\frac{1}{2}}[\frac{\exp(-iE_i^{\prime}/kT)\sigma_s^{i+1}}{(i+2)\sigma_i\Theta_c}]^{\frac{1}{2i+4}}\Gamma^{\frac{i}{2i+4}}. \label{eqA6}
	\end{equation}
	
	\section{B. The nucleation density in experiments}
	A great deal of researchs have focused on the control of nucleation density in experiments. The nucleation density results and the growth conditions in experiments are listed in Table. \ref{S1}.
\setlength{\tabcolsep}{2pt}
\linespread{0.85}
\begin{longtable*}{p{2.1cm}<{\centering} p{1.8cm}<{\centering} p{5.9cm}<{\centering} c p{3.5cm}<{\centering} p{0.5cm}<{\centering} }
	
	\caption{Summary of CVD growth conditions and nucleation density results reported in the literatures to control nucleation density in graphene growth.}\label{S1}\\
	
	\hline \hline
	
	\thead{Growth\\ technique} & \thead{Temperature\\($\rm ^\circ C $) }& \thead{Growth\\ condition} & \thead{Nucleation\\ density (mm$^{-2}$)} & \thead{Nucleation \\control method} & Ref.\\	\hline
	\endfirsthead
	
	\thead{Cu foil\\APCVD} & 1050 & \thead{15 sccm CH$_4$\\(500 ppm in Ar)}& 0.1 - 1 &\thead{ foil slightly\\ oxidizing} & \cite{Ganlin-2013} \\ \hline
	
	\thead{Cu foil\\APCVD} & 950-1080 & \thead{10.5 mTorr CH$_4$ (0.1$\%$ in Ar)\\ 19Torr H$_2$ (2.5$\%$ in Ar) } & 10$^4$ - 10$^7$ & changing $T$ & \cite{Vlassiouk-2013} \\ \hline
	
	\thead{O$_2$-assisted\\ Cu pocket\\ LPCVD} & 1035 & \thead{$P$(CH$_4$) = 0.001 Torr\\$P$(H$_2$) = 0.1 Torr } & 1 - 10$^3$  & O$_2$ exposing time &  \protect\cite{Haoyufeng-2013}\\ \hline
	
	\thead{Cu foil\\LPCVD} & 950-1050 & \thead{730 sccm CH$_4$ (100 ppm in Ar)\\130 sccm H$_2$. $P$(total) = 5 Torr} & 10$^4$ - 10$^6$  & changing $T$ & \cite{Vlassiouk-2013}\\ \hline
	
	\thead{Cu foil\\LPCVD} & 950-1050 & \thead{0.3 sccm CH$_4$, 15 sccm H$_2$\\ $P$(total) = 0.2 Torr} & 10$^2$ - 10$^3$  & changing $T$ & \cite{Vlassiouk-2013}\\ \hline
	
	\thead{Cu foil\\LPCVD} & 1070 & \thead{H$_2$/CH$_4$ = 1320\\ $P$(total) = 1-1000 mbar} & 10$^3$ - 5$\times$10$^3$  & \thead{changing reactor\\ pressure} & \cite{Zhouhailong-2013}\\ \hline
	
	\thead{Cu foil\\LPCVD} & 1070 & \thead{0.5 sccm CH$_4$, 10 sccm H$_2$ \\ $P$(total) = 1 Torr} & 10 - 10$^3$  & \thead{ foil slightly\\ oxidizing} & \cite{Eres-2014}\\ \hline	
	
	\thead{Cu foil\\LPCVD} & 1065 & \thead{Ar/H$_2$/(0.1$\%$)CH$_4$ = 250/26/7  \\ $P$(total) = 50 mbar} & 10 - 10$^2$  & \thead{ foil electropolishing} & \cite{Braeuninger-2016}\\ \hline
	
	\thead{Cu foil\\LPCVD} & 1065 & \thead{Ar/H$_2$/(0.1$\%$)CH$_4$ = 250/26/7  \\ $P$(total) = 50 mbar} & 10$^{-2}$ - 10  & \thead{ backside oxidizing} & \cite{Braeuninger-2016}\\ \hline
	
	\thead{Cu foil\\LPCVD} & 1065 & \thead{Ar/H$_2$/(0.1$\%$)CH$_4$ = 250/26/7  \\ $P$(total) = 50 mbar} & 10$^2$ - 10$^4$  & \thead{ foil surface etching} & \cite{Braeuninger-2016}\\ \hline
	\thead{Cu foil\\LPCVD} & 1040 & \thead{5 sccm CH$_4$ 20 sccm H$_2$  \\ $P$(total) = 12.5 - 20 Pa} & 10$^2$ - 10$^4$  & \thead{changing reactor\\ pressure} & \cite{Chaitoglou-2016}\\ \hline
	
	\thead{Cu(100) foil\\LPCVD} & 1020 & \thead{1 sccm CH$_4$ 20-1700 sccm H$_2$}   & 10$^{-6}$ - 10$^{-2}$  & \thead{changing H$_2$\\ flow rate} & \cite{Sunluzhao-2016}\\ \hline	
	
	\thead{CuNi alloy\\APCVD} & 1065 & \thead{Ar/H$_2$/(0.5$\%$)CH$_4$ = 300/15/60-150  \\ $P$(total) = 1 atm} & 10$^{-6}$ - 10$^{-4}$  & \thead{ changing CH$_4$\\ flow rate} & \cite{Wutianru-2016}\\ \hline
	
	\thead{Cu foil\\APCVD} & 1065 & \thead{Ar/H$_2$/(0.5$\%$)CH$_4$ = 300/15/15-45  \\ $P$(total) = 1 atm} & 10$^{-3}$ - 10$^{-2}$  & \thead{ changing CH$_4$\\ flow rate} & \cite{Wutianru-2016}\\ \hline
	
	\thead{O$_2$-assisted\\ Cu pocket\\ LPCVD} & 1060 & \thead{0.3 sccm CH$_4$  \\ $P$(H$_2$) = 10 - 870 Pa} & 1 - 10 & \thead{ changing H$_2$\\ pressure} & \cite{Zhaopei-2016}\\ \hline
	
	\thead{ Cu foil\\ LPCVD} & 970-1070 & \thead{5 sccm CH$_4$, 20 sccm H$_2$  \\ $P$(total) = 15 Pa} & 10$^2$ - 10$^3$ & \thead{ changing $T$} & \cite{Chaitoglou-2017}\\ \hline
	
	\thead{Cu foil\\ LPCVD} & 1040 & \thead{Ar/H$_2$/(0.9$\%$)CH$_4$ = 320/10/1.6 sccm \\ 0.2 - 4 sccm (0.9$\%$)O$_2$} & 10$^{-3}$ - 10$^{-2}$ & \thead{ changing O$_2$\\ flow rate} & \cite{Guowei-2016}\\ \hline
	
	\thead{ CuNi alloy\\ APCVD} & 1075 & \thead{Ar/H$_2$/(1$\%$)CH$_4$ =\\ 500/50/5 sccm} & 1 - 10 & \thead{ changing Ni content\\ in substrate} &\cite{Huangming-2018}\\ \hline
	
	\thead{Cu foil\\ LPCVD} & 1035-1075 & \thead{H$_2$/(0.9$\%$)CH$_4$ = 120 \\  $P$(total) = 10 Torr} & 10 - 10$^{2}$ & \thead{ changing $T$ } & \cite{Chanchaiachung-2018}\\ \hline
	
	\thead{ CuNi alloy\\ APCVD} & 1050 & \thead{ Ar/H$_2$/(0.5$\%$)CH$_4$ =\\ 300/15/60-150 sccm}   & 10$^{-6}$ - 10$^{-4}$ & \thead{ changing CH$_4$\\ flow rate} & \cite{Liuyifan-2018}\\ \hline
	
	\thead{ Co np Cu foil\\ APCVD} & 1030 & \thead{1-10 sccm CH$_4$ (22.5-112.5 mTorr)\\10 sccm H$_2$ (116.3 mTorr)}   & 10$^{3}$ - 10$^{4}$ & \thead{ changing \\CH$_4$/H$_2$ ratio} & \cite{Yinghao-2019}\\ 
	
	\hline\hline
\end{longtable*}
\linespread{1.5}

\end{document}